%Paper: hep-th/9312154
%From: David Gershon <GERSHON@taunivm.tau.ac.il>
%Date: Fri, 17 Dec 93 19:16:11 IST

% The following is a TeX file
%macropackage=phyzzx
\overfullrule=0pt
\font\boldm=cmbsy10
\font\boldmm=cmb10

\def\J{\hbox{\boldm\char'112}}
\def\T{\hbox{\boldmm\char'124}}
\font\boldmo=cmr10
\def\oe{\hbox{\boldmo\char'111}}
\font\boldmm=cmfi10

\def\Z{\hbox{\boldmm\char'132}}
\line{\hfill TAUP-2129-93}
\vskip .1in
\line{\hfill Dec,1993}
\date{}
\vskip .5 true cm
\titlepage
\title{\bf  EXACT O(d,d) TRANSFORMATIONS IN WZW MODELS}
\vskip 1 true cm
\centerline{\caps ~DAVID ~GERSHON}\footnote{1}{Work supported in
part by the US-Israel Binational
Science Foundation and the Israeli Academy of Sciences.}
\centerline{\it School of Physics and Astronomy}
\centerline{\it Beverly and Raymond Sackler Faculty of Exact Sciences}
\centerline{\it Tel-Aviv University, Tel-Aviv 69978, Israel$\dagger$}
\footnote{\dagger}{e-mail: GERSHON@TAUNIVM.BITNET}
\vskip 2 true cm
\abstract
Using the algebraic Hamiltonian approach, we derive the exact to
all orders O(d,d) transformations of the metric and the dilaton
field in WZW  and WZW coset models for both compact and non-compact
groups. It is shown that  under the
exact $O(d)\times O(d)$  transformation only the leading order
of the inverse metric $G^{-1}$ is transformed.
%shown that when writing the exact inverse metric as the one loop order
%part plus all the $\O({1\over k})$ contribution, under the
%part is transformed while the $\O({1\over k})$ part is unchanged.
The quantity  $\sqrt{G}\exp(\Phi) $ is
the same in all the dual models and in particular is independent of
$k$. We also show that the exact metric and dilaton field
 that correspond to $G/U(1)^d$ WZW can be obtained by applying
  the exact $O(d,d)$ transformations on the (ungauged) WZW, a result
 that was known to one loop order only.
 As an example we give the $O(2,2)$
transformations in the $SL(2,R)$ WZW that transform
 to its dual exact models. These include also the exact
3D black string and the exact 2D black hole
with an extra $U(1)$ free field.

\endpage
\chapter{\it Introduction}

$O(d,d)$ symmetries were originally discovered for toroidal
compactification of closed string theories\Ref\toroidal{K. Kikkawa
and M. Yamasaki\journal Phys.Lett.&B149 (84) 357;\nextline
N. Sakai and I. Senda\journal Prog.Teor.Phys.Suppl.&75 (86)692;
\nextline
K. Narain, M.Sarmadi and E. Wittem\journal Nucl.Phys.&B279 (87)369}
as a generalization of the familiar $R\rightarrow 1/R$ symmetry in
conformal field theory (CFT).
The global structure of the moduli space is rather interesting:
points corresponding to different backgrounds are related by duality
group isomorphic to $O(d,d,\Z)$ of discrete symmetries\Ref\malkin{A.
Giveon, N. Malkin and E. Rabinovici\journal Phys.Lett.&B220 (89)
551; \nextline
A. Shapere and F. Wilzcek\journal Nucl.Phys. &B320 (89) 669;\nextline
A. Giveon, E. Rabinovici and G. Veneziano\journal Nucl.Phys.&B322 (89)
 167}.

These duality transformations are generalized to string theories with
curved background.
It was shown by
Buscher\Ref\buscher{T. Buscher\journal Phys.Lett. & B201 (88)466}
 that duality symmetry of conformal backgrounds with
one isometry
transforms from one CFT to another (to one loop oredr). Subsequently,
it was  noticed\Ref\veneziano{K. Meissner and
G. Veneziano\journal Phys.Lett.&B267(91)33} that if we start with
the low energy effective action in string theory and restrict to space
translation invariant but time dependent background, the effective
action exhibits an $O(d,d)$ symmetry.
The generalization of the $O(d,d)$ symmetry in curved backgrounds
with $d$ isometries\Ref\sen{A. Sen \journal
Phys.Lett. &B271 (91) 295} and in the Heterotic
strings\Ref\hassan{S.F. Hassan and A. Sen\journal Nucl.Phys.&B375 (92)
103}  was
proven  to be an exact symmetry of string theories by
means of string field theory. The corresponding $O(d,d)$ transformations
could be derived to one loop order in $\alpha '$ only.
Giveon and Rocek\Ref\amitg{A. Giveon and
M. Rocek\journal Nucl.Phy. &B380 (92) 128} have used the $\sigma$-model
approach and showed the one loop order $O(d,d)$ transformations for
a general CFT background with $d$ isometries.  Duality symmetries
from non-abelian isometries were considered later\Ref\quevedo{X.C. de
la Ossa and F. Quevedo, "Duality Symmetries from Non-Abelian Isometries
in String Theory", preprint NEIP92-004} where it was found that
this new duality transformation maps spaces with non-abelian duality
to spaces that may have no isometries at all. Such symmetries were
also discussed in\Ref\gaume{E. Alvarez, L. Alvarez-Gaume, J.L.F.
Barbon and Y. Lozano, "Some Global Aspects of Duality in String Theory"
, preprint CERN-TH.6991/93}.

$O(d,d)$ symmetries are powerful and intriguing.
Beside providing a better understanding of the moduli space of a
given solution, $O(d,d)$ symmetry   leads to striking cosmological
consequences which we do not fully understand,
 such as those discussed in\Ref\vafa{R. Brandenberger and
C. Vafa\journal Nucl.Phys.&B316 (89) 391}\Ref\giveon{A. Giveon\journal
Mod.Phys.Lett.&A6 (91) 2843}
\Ref\amit{A. Giveon and E. Kiritsis, "Axial-Vector Duality as a Gauged
Symmetry and Topology Change in String Theory",
 preprint CERN-TH.6816/93}\Ref\gasperini{M. Gasperini, J. Maharana
and G. Veneziano \journal phys.Lett. &B296 (92) 51}.

The duality symmetries of WZW and WZW coset models were discussed
in\Ref\elias{E. Kiritsis\journal Nucl.Phys.&B405 (93) 109}.
 The exact underlying symmetry responsible
for semiclassical duality was identified with the symmetry under
affine Weyl transformations. This identification shows that in the
compact and unitary case they are exact symmetries of string theory
to all orders in $\alpha '$. Duality transformations were shown
\Ref\ashoke{S.F. Hassan and A. Sen\journal Nucl.Phys. &B405 (93) 143}
\Ref\chiara{M. Hanningson and C. Nappi, "Duality, Marginal
Perturbations and Gaugings", preprint IASSNS-HEP-92-88}
 to be equivalent to
integrated marginal perturbations by bilinears in the chiral currents.

In a previous paper\Ref\gershon{
D. Gershon, "Semiclassical vs. Exact Solutions of Charged Black
Holes in Four Dimensions and Exact $O(d,d)$ duality, preprint TAUP-2121}
 we have
shown the exact to all orders $O(d,d)$ transformations of the metric
and the dilaton field for   WZW coset models with d isometries
which have a  certain type of  a background.
The aim of this paper is to generalize this result
to  general WZW and WZW coset models with abelian isometries.
\footnote\dagger{In the case of
$SL(2,R)/U(1)$ and $SU(2)/U(1)$ there is a regularization scheme
where the semiclassical background receives no higher order
corrections in $\alpha'$\Ref\tsin{A.A. Tseytlin \journal Phys.Lett.
&B317 (93) 559}. In such a case the semiclassical
$O(d,d)$ symmetry transformations are  also exact
$\lbrack\amit\rbrack$.}

 The paper is organized as follows. In section 2 we just give the
expressions for the $O(d,d)$ transformations in one loop order for
the general case with $d$ isometries. In section 3 we derive an
expression for the exact metric and dilaton field for $G/U(1)^d$
WZW coset models by using the algebraic Hamiltonian approach.
In section 4 we derive the exact $O(d)\times O(d)$ dual models
of the coset models in section 3 and compare with the one loop
 order transformations. In section 5 we  derive the exact dual
 models to ungauged WZW models. In particular we show that
 the exact coset models can be obtained by exact duality from the
 ungauged WZW. In section 6 we demonstrate the exact duality
  transformations in the case of $SL(2,R)$ WZW and get also familiar
 models, such as the exact 3D black string. In section 7 we
 derive the transformations in the case of non-compact groups where
 $n$ isometries correspond to compact coordinates and $d-n$
 correspond to non-compact coordinates. The duality in this case
 is generated by $O(n,d-n)\times O(n,d-n)$ (beside constant coordinate
 transformations and a constant shift of the antisymmetric tensor).
 Section 8 is reserved for summary and remarks.

\chapter{\it $O(d,d)$ transformations in one loop order}

Consider a general $\sigma$-model with $d$ isometries
that correspond to a CFT
$$S={k\over 2\pi}\int d^2\sigma ( F_{\mu\nu}(X)
\partial_+X^{\mu}\partial_-
X^{\nu} +E_{\mu i}(X)\partial_-
Y^i\partial_+X^{\mu}
+\tilde E_{i\mu}(X)
\partial_+Y^i\partial_-X^{\mu}
$$$$+D_{ij}(X)\partial_+Y^i\partial_-Y^j)
-{1\over 8\pi}\int d^2\sigma\sqrt{h} R^{(2)}\Phi(X)\eqn\aa$$
where the background has $D$ target space dimensions with
$\mu=1,...,D-d$ and $i=1,...,d$.
$\sqrt{h}$ and  $R^{(2)}$ are
the determinant of the metric and the curvature  in the Riemann
surface, respectively,
and $\Phi$ is the dilaton field. This background corresponds to
the target-space metric $G$ and the antisymmetric tensor $B$ with
 $$G={1\over 2}\left (\matrix{(F+F^T)&
E+\tilde E^T\cr E^T+\tilde E&D+D^T\cr}\right )\;\;\;\;
B={1\over 2}\left (\matrix{(F-F^T)&
E-\tilde E^T\cr -E^T+\tilde E&D-D^T\cr}\right )\eqn\aaa$$
 The background \aa\
exhibits the $O(d,d)$ symmetry whose generators correspond to the
following symmetry transformations$\lbrack\amitg\rbrack$:
\nextline
(a). $O(d)\times O(d)$ symmetry transformations, under which
$$D\rightarrow D'=
\lbrack (O_1+O_2)D+(O_1-O_2)\rbrack\lbrack (O_1-O_2)D
+(O_1+O_2)\rbrack^{-1}\eqn\ab$$
$$E\rightarrow E'={1\over 2}
\lbrack (O_1+O_2)-D'(O_1-O_2)\rbrack^{-1}E\eqn\ad$$
$$\tilde E\rightarrow \tilde E'=2\tilde
E\lbrack (O_1-O_2)D+(O_1+O_2)\rbrack^{-1}\eqn\c$$
$$F\rightarrow F'=F-\tilde E
\lbrack (O_1-O_2)D+(O_1+O_2)\rbrack^{-1}(O_1-O_2)
E\eqn\ae$$
$$\Phi\rightarrow \Phi'=
\Phi+{1\over 2}\ln ({\det G\over \det G'})\eqn\af$$
where  $G'$ is the
 transformed target space metric. In the above
$O_1$ and $O_2$ are two constant $O(d)$ matrices.
These symmetry transformations can be derived by gauging a $U(1)^d$
subgroup in a "larger" background with $D+2d$ target space dimensions
with $2d$ isometries$\lbrack\amitg\rbrack$
and are correct to one loop order only.\nextline
(b). Coordinate transformations: $Y^i\rightarrow A^i_j Y^j$ where
$A$ is a $GL(d,R)$ constant matrix. The transformations with
$AA^T=\oe$ are already included in (a).\nextline
(c). A constant shift of the antisymmetric tensor $B_{ij}\rightarrow
B_{ij}+C_{ij}$ where $C$ is a $d\times d$ antisymmetric matrix.\nextline
Notice that the transformations in (b) and (c) are exact to all orders
in $\alpha'$ since the equations of motions to all orders are covariant
and  depend only
on the torsion $H_{\alpha\beta\gamma}$.

\chapter{\it  Exact metric and dilaton in WZW coset models}

In order to find the exact to all orders
metric and dilaton field we shall be
using  the algebraic Hamiltonian approach for $G/H$ coset models.
This approach was derived in\Ref\verlinde{R.
   Dijgraaf, E. Verlinde and H. Verlinde\journal
Nucl.Phys&B371 (92) 269 }
and systematized in\Ref\sfetsos{I. Bars and K. Sfetsos
 \journal Phys.Rev. &D46 (92) 4510}.\footnote{\dagger}{In the
 $SL(2,R)/U(1)$ model it was shown
\Ref\tseyt{A.A. Tseytlin\journal Phys.Lett.&B268 (91)175}\Ref\jack{
I. Jack, D.R.T. Jones and P. Panvel \journal Nucl.Phys.&B393 (93) 95}
that the background obtained by
this method satisfies the $\beta$-function equations at least up to
the fourth order in $\alpha '$. In the $SL(2,R)\times U(1)/U(1)$
coset it was shown\Ref\tsesfet{K. Sfetsos and A.A. Tseytlin,
"Antisymmetric Tensor Coupling  and Conformal Invariance in Sigma
Models Corresponding to Gauged WZNW Theories",
preprint CERN-TH.6969/93} that the exact background satisfies the
$\beta$-function equations at least up to the second order.}
 However,  this method is  not helpful in
calculating the exact antisymmetric tensor. Therefore we shall not
discuss the exact transformations of the antisymmetric tensor in this
paper.

Let us first briefly describe the method for coset models $G/H$.
  (For a review see \Ref\itz{I. Bars,
"Curved Spaces Geometry for Strings and Affine Non-Compact Algebras",
preprint USC-93/HEP-B3, to appear in {\it Interface Between
Mathematics and Physics}, Ed. S. T. Yau}.)
Denote by $J^G_a$ ($a=1,...,\dim G$)
and $J^H_i$  ($i=1,...,dim H$)
the
currents of the group $G$ and its subgroup $H$, respectively
 and $J^G_{a,n}, J^H_{i,n}$ are their
``Fourier" components in the Kac-Moody algebra.
 $L_0$ is the zero generator of the Virasoro
algebra. Then the ground state $T$ (the Tachyon) satisfies the
following conditions:
$$(L_0+\bar L_0-2)T=0\eqn\mmm$$$$(J^H_0+\bar J^H_0)T=0\eqn\mmb$$$$
J^G_nT=\bar J^G_nT=0,\;\;n\ge 1\eqn\mma$$
Here$$L_0={\Delta_G\over k-\tilde c_G}
-{\Delta_H\over k-\tilde c_H}\;\;\;\;\;\;\;\;\;\;\;\;\;\;
\bar L_0={\bar \Delta_G\over k-\tilde c_G}-
{\bar\Delta_H\over k-\tilde c_H}\eqn\ghp$$
and $\Delta_G$,$\Delta_H$ are the Casimir operators in $G$ and in $H$,
\ie,  $\Delta_G=J^G\cdot J^G$, $\bar\Delta_G={\bar J}^G\cdot{\bar J}^G$,
$\Delta_H=J^H\cdot J^H$, $\bar \Delta_H={\bar J}^H\cdot{\bar J}^H$,
 and $\tilde c_G,\tilde c_H$ are the coexter of $G,H$ respectively.
 In the language of the group elements $g$ and the left and right
 gauged subgroup elements $h_L$ and $h_R$,
  the condition \mmb\
is a remnant of the gauge invariance
$T(h_L g h_R^{-1}) =T(g)$ which demands that the tachyon is a singlet
under the action of the subgroup $H$.
Now we parameterize the group elements of $G$ with the coordinates
$X_{\mu}$, $\mu=1,...,N=\dim G$ and express the currents in terms of
first order differential operators of $X_{\mu}$ which satisfy the
Lie algebra of the group (so that
$L$ and $L_0$ become a differential operator). Finally we
define gauge invariant coordinates $\tilde X_{\mu}$, $\mu=1,...,D=
\dim G-\dim H$ so that when the tachyon is $T=T(\tilde X)$ it
automatically satisfies the condition \mmb.
Now we write the Casimir operators in terms of $\tilde X_{\mu}$,
by simply using the chain rule.
As is well known, the effective action for the Tachyon is
$$S(T)=\int d^DX\sqrt{-G}e^{\Phi}(G^{\mu\nu}\partial_{\mu}T\partial_{\nu}
T-V(T))\eqn\ert$$ where $\Phi$ is the dilaton field and $V(T)$ is the
Tachyon potential. On the other hand, since the Tachyon is completely
defined through the action of the zero modes, its action is equivalent
to $$S(T)=\int d^DX\sqrt{-G}e^{\Phi}(T(L_0+\bar L_0)T-V(T))\eqn\hami$$
 From the action  \ert \ and the action
 \hami, expressed in terms of $\tilde X_{\mu}$, we obtain
$$L_0+\bar L_0=-e^{-\Phi}{1\over \sqrt{-G}}\partial_{\mu}(e^{\Phi}
\sqrt{-G}G^{\mu\nu}\partial_{\nu})\eqn\qqqq$$ from which we find the
exact metric and the exact dilaton field.

This approach agrees with the fact that (ungauged) WZW
are exact to all orders in ${1\over k}$ up to a shift of the level
$k\rightarrow k-\tilde c_G$.
To get the one loop order background one should put
$\tilde c_G=\tilde c_H=0$ in
$L_0+\bar L_0$ of the gauged model, which is equivalent
to taking astronomically large $k$ and neglecting
$\tilde c_G$ and $\tilde c_H$.

Now, consider a general $D+d$ dimensional terget space that
correspond to a (ungauged) WZW model based on a group $G$ with level
$k$ which has a subgroup $U(1)^d$.
  We pick a basis ${T^i,\;i=1,...,d}$ in the Cartan subalgebra
  with $\lbrack T^i,T^j\rbrack=0$ and $\tr T^iT^j=\delta^{ij}$.
 Now we parameterize the group elements as
$$g=e^{i\sum_{i=1}^d\theta_1^i T^i}\tilde g(X) e^{i\sum_{i=1}^d
  \theta_2^i T^i}\eqn\aaae$$
 so the action can be written as
$$S={k\over 2\pi}\int d^2\sigma(\partial_+\theta_1^i\partial_-
\theta_1^i+\partial_+\theta_2^i\partial_-\theta_2^i+2M_{ij}(X)
\partial_- \theta_1^i\partial_+\theta_2^j+2N^1_{\mu i}(X)
\partial_+X^{\mu}\partial_-\theta_1^i$$$$+2N^2_{\mu i}(X)
\partial_-X^{\mu}
\partial_+\theta_2^i
+F_{\mu\nu}(X)\partial_+X^{\mu}\partial_- X^{\nu})\eqn\bb$$
with $\mu=1,...,D-d$ and $i=1,...,d$.
This action \bb\ is invariant under $U(1)_L^d\times U(1)_R^d$ symmetry
generated by the conserved holomorphic currents that correspond to
translational symmetry in $\theta_1^i$ and $\theta_2^i$
$$J^i=\partial_+\theta_1^i+M_{ij}\partial_+\theta_2^j
+N^1_{\mu i}\partial_+X^{\mu}\eqn\bba$$
$${\bar J}^i=\partial_-\theta_2^i+M_{ji}\partial_-\theta_1^j
+N^2_{\mu i}\partial_-X^{\mu}\eqn\bbaa$$
Now we want to describe the coset model $G/U(1)^d$ in the algebraic
Hamiltonian approach. The group $G$ is parameterized so that the
left and the right $U(1)$
currents are the commuting diffrential operators
$$J_j=i\partial_{\theta_1^j}\;\;\;;\;\; \bar J_j=i\partial_{\theta_2^j}
\;\;\;\;\;\;\;\;j=1,...,d\eqn\bbc$$
%The ungauged WZW action is exact to all orders
%up to a shift $k\rightarrow k-\tilde c_G$  (and $\Delta=\bar \Delta$).
We can choose the rest of the generators to correspond to the action
\bb (since the WZW is exact). Now we gauge the axial gauge.
In the gauged model
$$L_0+\bar L_0=-{1\over k-\tilde c_G}(\Sigma^{\mu}(X)\partial_{X^{\mu}}+
\Gamma_1^i(X)\partial_{\theta_1^i}+\Gamma_2^i(X)
\partial_{\theta_2^i}$$$$
+G^{\mu\nu}(X)
\partial_{X^{\mu}}\partial_{X^{\nu}}+2G_1^{\mu i}(X)
\partial_{X^{\mu}}\partial_{\theta_1^i}+2G_2^{\mu i}(X)
\partial_{X^{\mu}}
\partial_{\theta_2^i}$$$$
+L_1^{ij}(X)\partial_{\theta_1^i}\partial_{\theta_1^j}
+L_2^{ij}(X)\partial_{\theta_2^i}\partial_{\theta_2^j}
+2P^{ij}(X)
\partial_{\theta_1^i}\partial_{\theta_2^j}) +{1\over k}\sum_{i=1}
^d(\partial^2_{\theta_1^i}+\partial^2_{\theta_2^i})\eqn\cc$$
where $G^{\mu\nu},G_1^{\mu i},G_2^{\mu i},L_1^{ij},L_2^{ij}$ and $P^{ij}
$ are the components of the inverse of the metric in the WZW action, \ie,
$$G^{-1}=\left (\matrix{ G&G_1&G_2\cr G_1^T&L_1&P\cr G_2^T&P^T&L_2\cr}
\right ) =\left (\matrix{{1\over 2}(F+F^T)
&N^1&N^2\cr {N^1}^T&\oe&M\cr {N_2}^T&M^T&\oe\cr}\right )^{-1}
\eqn\ccddd$$  and $\Sigma,
\Gamma_1,\Gamma_2$ are obtained from \qqqq\ with $\Phi=0$.
(notice that we have written all matrices with upper indices for
convenience. This  has nothing with covariant raising indices.)

Now we should define $d$ gauge
invariant coordinates which we denote by $Y^i$ and
in addition to $X^{\mu}$, which are not transformed, the target space
of the gauged action is $D$ dimensional.
 In the axial gauge  the gauge invariant coordinates
should satisfy $d$ conditions $$(J_i+\bar J_i)Y^j=0\;\;;\;\;
i,j=1,...,d\eqn\ccca$$
Taking $$Y^i=\theta_1^i-\theta_2^i\eqn\cccb$$ we obtain
$$L_0+\bar L_0=-{1\over k-\tilde c_G}\lbrack
\Sigma_{\mu}(X)\partial_{X^{\mu}}+
(\Gamma^1_i-\Gamma^2_i)\partial_{Y^i}
+G^{\mu\nu}\partial_{X^{\mu}}\partial_{X^{\nu}}+2(G_1^{\mu i}
-G_2^{\mu i})\partial_{X^{\mu}}\partial_{Y^i}$$$$
+(L_1^{ij}+L_2^{ij}-2\delta^{ij} -2P^{ij})\partial_{Y^i}\partial_{Y^j}
+{2\tilde c_G\over k}\delta^{ij}\partial_{Y^i}\partial_{Y^j}
\rbrack\eqn\ddd$$Define
 $\tilde I$ to be the $D\times D$ matrix
$$\tilde I=\left (\matrix {0&0\cr 0&I\cr}\right ) \eqn\stam$$
and   $I$ is the $d\times d$ unit matrix.
{}From the  expression \ddd\ we see that
$$G^{-1}_{exact}={1\over k-\tilde  c_G}
(G^{-1}_{classical}+{2\tilde c_G\over k}\tilde I)\eqn\dda$$
 Here we omitted the pre-factor $k$ infront of $G_{classical}$.
We could, of course,
choose different gauge invariant coordinates $Y^i=C^i_j(\theta_1
^j-\theta_2^j)$, where $C$ is a constant $GL(d)$ matrix,
but this is simply a coordinate transformation of the
$Y^i$ coordinates. In this case
the ${1\over k}$ correction to the semiclassical
metric becomes $(C^TC){\tilde c_G\over k}$ in the $(ij)$ directions.
%(Notice that if the coordinates
%$\theta$ are compactified we should restrict to $GL(d,\Z)$ in order
%to preserve the periodicity, otherwise we obtain exact CFTs
%which are not equivalent to .)

\chapter{\it  Exact Duality Transformations}

 All the $O(d)\times O(d)$ dual models to the gauged model in the
previous section
 can be obtained by all the
 different anomaly free  gaugings (generators) of the $U(1)^d$
subgroup. The condition for the anomaly
cancellation is\Ref\witten{E. Witten
 \journal Commun.Math.Phys. &144 (92)189}
$$\tr T_L^aT_L^b=\tr T_R^a T_R^b \; \;\;\;a,b=1,...,\dim H\eqn\bbb$$
where $T_L^a,T_R^a$ are the generators of the left and right
gauged subgroups.
 Instead of looking for all the anomaly free gaugings we shall
use the following method, which is an application  of the
idea used in $\lbrack\amitg\rbrack$ for the $\sigma$- model action.
The ungauged action\bb, or equivalently, the Casimir operator of
the ungauged model, is invariant under the transformation
$$\theta_1\rightarrow \theta_1'=O_1\theta_1\;\;;\;\;
\theta_2\rightarrow \theta_2'=O_2\theta_2$$
$$L_1\rightarrow L_1'=O_1L_1O_1^T\;\;;\;\;
L_2\rightarrow L_2'=O_2L_2O_2^T\;\;;\;\;
P\rightarrow P'=O_1PO_2^T$$
$$G_1\rightarrow G'_1=G_1O_1^T\;\;;\;\;G_2\rightarrow G'_2=G_2O_2^T$$
$$\Gamma_1\rightarrow \Gamma_1'=\Gamma_1O_1^T\;\;;\;\;
\Gamma_2\rightarrow \Gamma_2'=\Gamma_2O_2^T\eqn\dde$$
where $O_1,O_2$  are two $O(d)$ matrices (or $O(d,\Z)$ if we
want to preserve periodicity).
Now we  write $L_0+\bar L_0$ in the WZW action with
respect to the transformed coordinates $\theta_1',\theta_2'$
and
gauge the $U(1)^d$ subgroup generated by the currents $J_i=i\partial
_{\theta_1'^i}$ and $\bar J_i=i\partial_{\theta_2'^i}$, which
is still an anomaly free gauging. Using $\tr T^iT^j=\delta^{ij}$ it
is easy to see that the new generators, which are linear combinations
of the $T^i$s satisfy the condition \bbb.
 The result (in the $\theta'$ coordinates) is
a rotation of the Casimir operator of the ungauged action while
the gauged subgroup in unchanged. Therefore we get an expression for
all the $O(d)\times O(d)$ dual models:\nextline
$G^{-1}$\nextline$$={1\over k-\tilde c_G}
\left (\matrix {G& G_1O_1^T-G_2O_2
^T\cr O_1 G^T_1-O_2G^T_2& O_1L_1O_1^T+O_2L_2O_2^T
-O_1PO_2^T-O_2P^TO_1^T-2\oe
+{2\tilde c_G\over k}\oe \cr}\right )\eqn\ggff$$
for two general  $O(d)$ matrices. The important point
to notice is that the ${1\over k}$ correction to the inverse metric
with respect to the semiclassical one is invariant under the
$O(d)\times O(d)$ transformations (also when it takes the form $C^TC
{\tilde c_G\over k}$).
 Finally, we recall that all the
one loop order metrics that are obtained by $O(d)\times O(d)$
transformations can also be obtained by different gaugings of the
ungauged WZW model. Therefore under the $O(d)\times O(d)$ transformations
the exact inverse metric transforms as
$$G^{-1}_{exact}\rightarrow {G'}^{-1}_{exact}={1\over k-\tilde c_G}
({G'}^{-1}_{classical}+{2\tilde c_G\over k}\tilde{\oe})\eqn\fffb$$
where $\tilde{\oe}$ is given in \stam \ and $G'_{classical}$ is the
semiclassical metric obtained after the one loop
 $O(d)\times O(d)$ transformations,
 as is written in \ab-\ae.
% Notice that in the general case the ${1\over k}$
%correction to the inverse metric is not necessary diagonal, but is
%always invariant under the $O(d)\times O(d)$ transformations.
Hence, we reach the following conclusion. Given an exact metric
that correspond to a coset model, in order to find all the
exact $O(d)\times O(d)$ dual models it is enough to know
the antisymmetric tensor to one loop order.

The transformation of the exact dilaton under the $O(d)\times O(d)$
can be derived as well. Since the matrix $G$ and the vector $\Sigma$
are invariant under the transformation, using \qqqq \
it is easy to see that the quantity
$\sqrt{G}e^{\Phi}$
is invariant under the transformation.
Therefore under the transformation $G_{exact}\rightarrow G'_{exact}$
we obtain $$\Phi_{exact}\rightarrow \Phi'_{exact}=\Phi_{exact}+{1\over 2}
\ln({\det G\over \det G'})\eqn\gg$$ where $G,G'$ correspond to the
exact metric and the transformed exact metric.

In particular, we can  construct the axial-vector duality. This
duality was revealed   in the $\sigma$-model
approach\Ref\kiritsis{E. Kiritsis
\journal Mod.Phys.Lett. &A6 (91) 2871}
and was proven to
be an exact symmetry of the CFT$\lbrack \amit\rbrack$.
If we take $O_1=-O_2=\oe$ we interchange the axial and vector gaugings.

\chapter{\it Exact $O(d,d)$ Transformations in (Ungauged) WZW Models}

The method described above can be used to construct the exact
$O(d,d)$ transformations also in ungauged WZW models. Consider
a WZW model based on a group $G$ which has a $U(1)^d$ subgroup.
This means that the sigma model has $2d$ isometries. Although the
ungauged action is exact to all orders up to a shift of the level
(which is just a pre-factor), the exact $O(d)\times O(d)$ transformations
introduce  ${1\over k}$ corrections in the dual models.
To apply the duality transformations we should consider an equivalent
model- $G\times U(1)_k^{2d}/U(1)_k^{2d}$.
We use the notations in \cc\  for the (ungauged) WZW model
and  parameterize the extra  $U(1)$
generators by the differential operators
 $i\partial_{\varphi_1^i}$ and $i\partial_{\varphi_2^i}$ taken with
level $k$.
This introduces the term $-{2\over k}
\sum_{i=1}^d(\partial^2_{\varphi_1^i}
+\partial^2_{\varphi_2^i})$ to $L_0+\bar L_0$ of
the ungauged model. Now we gauge
the axial $U(1)_L^{2d}\times U(1)_R^{2d}$ subgroup generated by
$$\J_i=
i\partial_{\theta_1^i}\;{\rm for}\;\; i=1,...,d\;\;; \J_i=
i\partial_{\varphi_1^{i-d}}\;
 {\rm for}\;\; i=d+1,...,2d$$
$$\bar{\J}_i=
i\partial_{\theta_2^i}\;{\rm for}\;\; i=1,...,d\;\;;\bar{\J}_i=
i\partial_{\varphi_2^{i-d}}\;{\rm for}\;\; i=d+1,...,2d
\eqn\ggaa$$
%so that the central charge of $\J_i$ and $\bar{\J}_i$ is $k$.
Notice that the ${1\over k}$ is simply a redefinition of the $U(1)$
free fields but it ensures that the condition for
anomaly cancellation \bbb\ is satisfied after the rotation we shall make.
Thus,  we have now
$$L_0+\bar L_0=-{1\over k-\tilde c_G}(\Sigma^{\mu}(X)\partial_{X^{\mu}}+
\Gamma_1^i(X)\partial_{\theta_1^i}+\Gamma_2^i(X)
\partial_{\theta_2^i}
+G^{\mu\nu}\partial_{X^{\mu}}\partial_{X^{\nu}}+2G_1^{\mu i}
\partial_{X^{\mu}}\partial_{\theta_1^i}$$$$
+2G_2^{\mu i}\partial_{X^{\mu}}\partial_{\theta_2^i}
+L_1^{ij}\partial_{\theta_1^i}\partial_{\theta_1^j}
+L_2^{ij}\partial_{\theta_2^i}\partial_{\theta_2^j}
+2P^{ij}\partial_{\theta_1^i}\partial_{\theta_2^j})
-{2\over k}\sum_{i=1}^d(\partial^2_{\varphi_1^i}
+\partial^2_{\varphi_2^i})$$$$
 +{1\over k}\sum_{i=1}
^d(\partial^2_{\theta_1^i}+\partial^2_{\theta_2^i}+
\partial^2_{\varphi_1^i}+\partial^2_{\varphi_2^i})\eqn\hhcc$$
To obtain the metric we define the
gauge invariant coordinates
$$Y_1^i=\theta_1^i-\theta_2^i\;\;;\;\;
Y_2^i=\varphi_1^i-\varphi_2^i\;\;;i=1,...,d\eqn\hhdd$$
and substitute in $L_0+\bar L_0$.

Now, in order to see the $O(2d)\times O(2d)$ duality we define the $2d$
dimensional
vector $\phi_1$ with $\phi_1^i=\theta_1^i$ for $
i=1,...,d$ and $\phi_1^i=\varphi_1^{i-d}$ for $i=d+1,...,2d$ and
similarly we define the
vector $\phi_2$ with $\phi_2^i=\theta_2^i$ for
$i=1,...,d$ and $\phi_2^i=
\varphi_2^{i-d}$ for $i=d+1,...,2d$.
Now we rewrite $L_0+\bar L_0$ in \hhcc \
as follows:
$$L_0+\bar L_0=-{1\over k-\tilde c_G}(\Sigma^{\mu}(X)\partial_{X^{\mu}}+
\tilde\Gamma_1^i(X)\partial_{\phi_1^i}+\tilde\Gamma_2^i(X)
\partial_{\phi_2^i}$$$$
+G^{\mu\nu}\partial_{X^{\mu}}\partial_{X^{\nu}}+2\tilde G_1^{\mu i}
\partial_{X^{\mu}}\partial_{\phi_1^i}+
2\tilde G_2^{\mu i}\partial_{X^{\mu}}
\partial_{\phi_2^i}$$$$
+\tilde L_1^{ij}\partial_{\phi_1^i}\partial_{\phi_1^j}
+\tilde L_2^{ij}\partial_{\phi_2^i}\partial_{\phi_2^j}
+2\tilde P^{ij}\partial_{\phi_1^i}\partial_{\phi_2^j})
+{1\over k}\sum_{i=1}
^{2d}(\partial^2_{\phi_1^i}+\partial^2_{\phi_2^i})\eqn\cc$$
where $\tilde \Gamma_1,\tilde \Gamma_2$ are the $2d$ dimensional
vectors
$$\tilde \Gamma_1=(\Gamma_1,{\bf 0})\;\;;\;\;
\tilde \Gamma_2=(\Gamma_2,{\bf 0})\eqn\hhha$$
 $\tilde G_1,\tilde G_2$ are $(D-d)\times 2d$ matrices
$$\tilde G_1=\left (\matrix{ G_1& {\bf 0}\cr}\right )
\;\;;\;\;
\tilde G_2=\left (\matrix{ G_2& {\bf 0}\cr}\right )\eqn\hhhb$$
and $P,L_1,L_2$ are $2d\times 2d$ matrices
$$\tilde P=\left (\matrix{ P& {\bf 0}\cr{\bf 0}& {\bf 0}\cr}\right )
\;\;;\;\;\tilde L_1=
\left (\matrix{ L_1& {\bf 0}\cr{\bf 0}&2(
 1-{\tilde c_G\over k})\oe\cr}\right )
\;\;;\;\;\tilde L_2=\left (\matrix{L_2& {\bf 0}\cr{\bf 0}&2
(1-{\tilde c_G\over k})\oe\cr}\right )
\eqn\hhhc$$
Now we can repeat the steps in section 4.
Taking $O_1$ and $O_2$ to be  two $O(2d)$ matrices,
the Casimir operator of the ungauged model is invariant under the
$O(2d)\times O(2d)$ transformations
$$\phi_1\rightarrow \phi'_1=O_1\phi_1\;\;;\;\;
\phi_2\rightarrow \phi'_2=O_2\phi_2$$
$$\tilde L_1\rightarrow \tilde L_1'=O_1\tilde L_1O_1^T\;\;;\;\;
\tilde L_2\rightarrow \tilde L_2'=O_2\tilde L_2O_2^T\;\;;\;\;
\tilde P\rightarrow \tilde P'=O_1\tilde PO_2^T$$
$$\tilde G_1\rightarrow \tilde G'_1=\tilde G_1O_1^T
\;\;;\;\;\tilde G_2\rightarrow \tilde G'_2=\tilde G_2O_2^T$$
$$\tilde \Gamma_1\rightarrow \tilde \Gamma_1'=
\tilde \Gamma_1O_1^T\;\;;\;\;\tilde \Gamma_2\rightarrow
\tilde\Gamma_2'=\tilde \Gamma_2O_2^T\eqn\dde$$
So all the $O(2d)\times O(2d)$ dual models to the (ungauged)
WZW model have the following metrics:\nextline
$G^{-1}$\nextline $$={1\over k-\tilde c_G}
\left (\matrix {G& \tilde G_1O_1^T-
\tilde G_2O_2^T\cr O_1 \tilde G^T_1-O_2
\tilde G^T_2&  O_1\tilde L_1O_1^T+O_2\tilde L_2O_2^T
-O_1\tilde PO_2^T-O_2\tilde P^TO_1^T -2\oe
+{2\tilde c_G\over k}\oe \cr}\right )\eqn\dff$$
 Thus,  although the ungauged WZW model is exact
to all orders (up to a shift in $k$), dual models receive non-trivial
${1\over k}$ corrections with respect to the semiclassical backgrounds.
To get the dilaton fields we simply need to calculate
the determinant of the dual metric, divide by the determinant of the
metric in the WZW action and take a log. Since
$\sqrt{G}e^{\Phi}$ is the same in all the dual models as
in the ungauged WZW, this quantity is independent of $k$ to all orders.
In particular, the coset models with $d$ extra free $U(1)$ fields are
dual to the ungauged WZW and therefore this property holds also for
the coset models. The coset models that were derive in section 3
are obtained by taking
 $O_1=O_2=I$ for the axial gauge and $O_1=-O_2=I$ for the vector
gauge. If we take
 $$O_1=I\;\;\;;\;\;\;O_2=-\left (\matrix {{\bf 0}& \oe\cr \oe&{\bf 0}\cr}
 \right )\eqn\sof$$we obtain the metric of the ungauged WZW in \ccddd.
 Thus we have produced a result known to one loop order
\Ref\kumar{A. Kumar \journal Phys.Lett. &B293 (92) 49}
\Ref\dudi{D, Gershon, "Coset Models Obtained by Twisting WZW Models
 and Stringy Charged Black Holes in 4D",
 preprint 2005-92, to appear in Phys.Rev.D},
  that
 the background that correspond to $G/U(1)^d$ WZW can be obtained
by applying $O(d,d)$ transformations on the ungauged WZW.

\chapter{\it Example: Exact Duality in $SL(2,R)$ WZW}

To demonstrate the above, we shall consider here the exact
dual models to the $SL(2,R)$ WZW. This model has raised some
interest recently with connection to the 3D black
hole\Ref\horo{G.T. Horowitz and D.L. Welch, "Exact Three Dimensional
Black Holes in String Theory",
 preprint NSF-ITP-92-21}.
In one loop order it is known
\Ref\asen{A. Sen \journal Phys.Lett.&B274 (92) 33}
that by duality
transformations it can be brought to the 3D black
string\Ref\horowit{J.H. Horne and G.T. Horowitzand\journal
Nucl.Phys. &B368 (92) 444;
 N. Ishibashi,
M. Li and A. Steif\journal Phys.Rev.Lett. &67 (91) 3336}
and to the 2D black hole with an extra free $U(1)$ field$\lbrack
 \kumar,\dudi \rbrack$.

In the case of $SL(2,R)$ WZW there are 2 isometries. We shall
demonstrate the $O(2)\times O(2)$ duality.
First let us parameterize the currents of the $SL(2,R)$ WZW as
$$J_3=i\partial_{\theta_L}\;\;\;\; ;\;\;\;\bar J_3=
i\partial_{\theta_R}$$
$$J_{\pm}=ie^{\pm \theta_L}({1\over 2}
\partial_r\pm {1\over \sinh 2r}(\partial_{\theta_R}
-\cosh 2r \partial_{\theta_L}))$$
$$\bar J_{\pm}=ie^{\pm \theta_R}({1\over 2}\partial_r\pm
{1\over \sinh 2r}(\partial_{\theta_L}
-\cosh 2r \partial_{\theta_R}))\eqn\qsl$$
Now we shall follow the procedure suggested in section 5. We introduce
2 additional $U(1)$ fields, denoted by $\varphi_L, \varphi_R$ and the
corresponding $U(1)$ currents (with level $k$) are $K=i\partial_{\varphi
_1}$ and $\bar K=i\partial_{\varphi_2}$. Next
we  gauge the currents $J_3,\bar J_3,K,\bar K$
Then in the model $SL(2,R)\times U(1)^2/U(1)^2$ we have
$$L_0+\bar L_0=-{2\over k-2}({1\over 4}
\partial^2_r+{1\over 2}\coth 2r \partial_r-
{1\over \sinh^22r}(\partial^2_{\theta_L}-2\cosh 2r\partial_{\theta_L}
\partial_{\theta_R}+\partial^2_{\theta_R})$$$$-{2\over k}(\partial^2
_{\varphi_1}+\partial^2_{\varphi_2})+{1\over k}(\partial^2_{\theta_L}
+\partial^2_{\theta_R}+\partial^2_{\varphi_1}+\partial^2_{\varphi_2})
\eqn\jjja$$ where we used $\tilde c_G=2$ for the $SL(2,R)$ group.
We  rewrite this expression as
$$L_0+\bar L_0=-{1\over k-2}({1\over 2}
\partial^2_r+\coth 2r \partial_r)
-\lbrack
(\partial_{\theta_L},\partial_{\varphi_1})A \left (\matrix{
\partial_{\theta_L}\cr\partial_{ \varphi_1}\cr}\right )
+(\partial_{\theta_L},\partial_{\varphi_1})B \left (\matrix{\partial_
{\theta_R}\cr \partial_{\varphi_2}\cr}\right )$$$$
+(\partial_{\theta_R},\partial_{\varphi_2})A \left (\matrix{\partial_{
\theta_R}\cr\partial_{ \varphi_1}\cr}\right )\rbrack
+{1\over k}( \partial^2_{\theta_L}+ \partial^2_{\theta_R}+
\partial^2_{\varphi_1}+\partial^2_{\varphi_2})  \eqn\jjc$$
where $$A=\left (\matrix {-{2\over (k-2)\sinh^22r}&0\cr 0&{2\over k}\cr}
\right )\;\;\;;\;\;\; B=\left (\matrix{{4\cosh 2r\over (k-2)\sinh^2 2r}
&0\cr 0&0\cr}\right ) \eqn\jjd$$
In the axial gauge we define the gauge invariant coordinates as
$$Y_1=\theta_L-\theta_R\;\;\;;\;\; Y_2=\varphi_1-\varphi_2\eqn\jjjkk$$
Then the dual models can be described with
$$L_0+\bar L_0=-{1\over k-2}({1\over 2}\partial^2_r+
\coth 2r \partial_r)
$$$$- (\partial_{Y_1},\partial_{Y_2})
(O_1 AO_1^T+O_2AO_2^T-O_1BO_2^T-{1\over k}\oe)\left (
\matrix{\partial_{Y_1}\cr\partial_{Y_2}\cr}\right )
\eqn\kkk$$ where $O_1,O_2$ are two $O(2)$ matrices.
Finally, let us take
$$O_1=\left (\matrix{\cos\alpha&\sin\alpha\cr -\sin\alpha&\cos\alpha\cr}
\right )\;\;;\;\; O_2=
\left (\matrix{\cos\beta &\sin\beta \cr -\sin\beta &\cos\beta \cr}\right
)\eqn\kkkaa$$ Substituting these matrices in \kkk, we obtain a general
expression for the inverse metric in the dual models:
%(we took a
%pre-factor ${1\over 2}$ in front of $G^{-1}$):
$$G^{rr}={1\over 2(k-2)}$$
$$G^{11}=-{2\over k-2}((\cos^2\alpha+\cos^2\beta){1\over \sinh^2 2r}
-(1-{2\over k})(\sin^2\alpha+\sin^2\beta-1)$$$$
+\cos\alpha\cos\beta{2\cosh 2r
\over \sinh^2 2r})$$
$$G^{22}=-{2\over k-2}((\sin^2\alpha+\sin^2\beta){1\over \sinh^2 2r}
-(1-{2\over k})(\cos^2\alpha+\cos^2\beta-1)$$$$
+\sin\alpha\sin\beta{2\cosh 2r
\over \sinh^2 2r})$$
$$G^{12}={2\over k-2}((\sin\alpha\cos\alpha+\sin\beta\cos\beta)
({1\over \sinh^2 2r}+1-{2\over k})+\sin(\alpha+\beta){\cosh 2r\over
\sinh^2 2r})\eqn\kkdd$$
 Let us now observe this metric.
For $\sin\beta =\cos\alpha=0$ we obtain the original ungauged WZW.
\nextline
For $\cos\alpha=\cos\beta=1$ we obtain the exact metric of the
2D black hole and an extra free $U(1)$ field (this is the
Lorentzian version of the solution in $\lbrack \verlinde\rbrack$).
Taking $t=2Y_1$ for the time and $X=2Y_1$ the metric with a pre-factor
2  is
$$dS^2=(k-2)dr^2-(k-2)(\coth^2r -{2\over k})^{-1}dt^2+
kdX^2$$$$     \Phi=\log(\sinh 2r)+{1\over 2}\log (\coth^2 r-{2\over k})
\eqn\llla$$
For $\cos\alpha=1$ and $\cos\beta=-1$ we obtain the dual metric of
the 2D black hole (that correspond to the vector gauging).
$$dS^2=(k-2)dr^2-(k-2)(\tanh^2r -{2\over k})^{-1}dt^2+
kdX^2$$$$     \Phi=\log(\sinh 2r)+{1\over 2}\log (\tanh^2 r-{2\over k})
\eqn\lllb$$
The exact background that correspond to the 3D black string
is obtained by a coordinate transformation on $Y_1,Y_2$.
We take $\cos\alpha=\sin\beta$ and define $t=Y_1-Y_2$ and
$X=Y_1+Y_2$. The metric we obtain is
$$dS^2=(k-2)dr^2-(k-2){\sinh^2 r\over 1+q+2q(1-{2\over k})\sinh^2 r}
dt^2$$$$
+(k-2){\cosh^2 r\over 1-q+2q(1-{2\over k})\cosh^2 r}dX^2$$$$
\Phi={1\over 2}\log\lbrack (1+q+2q(1-{2\over k})\sinh^2 r)(
 1-q+2q(1-{2\over k})\cosh^2 r)\rbrack
\eqn\lllc$$
where $q=\sin 2\alpha$. (These metric and dilaton field were
derived also in\Ref\sft{
K. Sfetsos\journal Nucl.Phys. &B389 (93) 424; I. Bars and K. Sfetsos
\journal Phys.Rev.&D48 (93) 844}.)
Notice that in all the models and in the ungauged WZW
$\sqrt{G}e^{\Phi}=\sinh 2r $.

\chapter{\it Exact $O(n,d-n)\times O(n,d-n)$ transformations}

In  the previous sections we were discussing WZW where the group
elements were parameterized by $g=
e^{i\sum_{i=1}^d\theta_1^i T^i}\tilde g(X) e^{i\sum_{i=1}^d
  \theta_2^i T^i}$. The results we got hold also when the
group elements are  parameterized by $$g
=e^{\sum_{i=1}^d\theta_1^i T^i}\tilde g(X) e^{\sum_{i=1}^d
  \theta_2^i T^i}\eqn\ssc$$
   (all the $\theta$ coordinates are non-compact)
  but requires a minus sign in front of the $(ij)$
components of the  exact metric in all the dual models.
In non-compact group we can also parameterize the group by
$$g=e^{i\sum_{i=1}^n\theta_1^i T^i+\sum_{i=n+1}^d \theta_1^i
T^i}\tilde g(X) e^{i\sum_{i=1}^n\theta_2^i T^i+\sum_{i=n+1}^d
\theta_2^i T^i}\eqn\new$$ ($2n$ compact and $2(d-n)$ non-compact fields)
where, as in section 3, $\tr T^iT^j=
\delta{ij}$.
In this case the ungauged action takes the form
$$S={k\over 2\pi}\int d^2\sigma(\sum_{i=1}^n(
\partial_+\theta_1^i\partial_-\theta_1^i
+\partial_+\theta_2^i\partial_-\theta_2^i)
-\sum_{i=n+1}^d( \partial_+ \theta_1^i\partial_-\theta_1^i+
\partial_+\theta_2^i\partial_-\theta_2^i)$$$$
+2M_{ij}(X)
\partial_- \theta_1^i\partial_+\theta_2^j
+2N^1_{\mu i}(X)
\partial_+X^{\mu}\partial_-\theta_1^i+2N^2_{\mu i}(X)
\partial_-X^{\mu}
\partial_+\theta_2^i$$$$
+F_{\mu\nu}(X)\partial_+X^{\mu}\partial_- X^{\nu})\eqn\ll$$
and the holomorphic currents $J^j=\tr \partial_+g g^{-1}T^j$ and
the anti-holomorphic currents  $\bar J^j=\tr g^{-1}\partial_-gT^j$
 with $j>n$ do not have a pre-factor $i$.
In the algebraic Hamiltonian approach we shall parameterize these
$U(1)$ currents as
$$J^j=i\partial_{\theta_1^j}\;\;{\rm for}\;\;i=1,...,n\;\;\;;\;\;\;
J^j=\partial_{\theta_1^j}\;\;{\rm for}\;\;i=n+1,...,d\eqn\gga$$
$$\bar J^j=i\partial_{\theta_2^j}\;\;{\rm for}\;\;i=1,...,n\;\;\;;\;\;\;
\bar J^j=\partial_{\theta_2^j}\;\;{\rm for}\;\;i=n+1,...,d\eqn\ggb$$
Then, in the gauged model we have
$$L_0+\bar L_0={\Delta_G+\bar \Delta_G\over k-\tilde c_G}
+{1\over k}\sum_{i=1}^n(\partial^2_{\theta_1^i}+\partial^2
_{\theta_2^i})
-{1\over k}\sum_{i=n+1}^d(\partial^2_{\theta_1^i}+\partial^2
_{\theta_2^i})$$$$=
{2\Delta_G\over k-\tilde c_G}
+{1\over k}\eta  (\partial^2_{\theta_1}+\partial^2_{\theta_2})
\eqn\gggc$$   where $\eta$ is the$d\times d$
matrix $\eta=diag(1,...,1,-1,...,-1)$ with $n$ entries 1.
(We have used $\Delta_G=\bar \Delta_G$.)

Let us consider now the rotation of the $\theta$ coordinates.
The translation of the coordinates $\theta_1^i,\theta_2^i$ with
$i=1,...,n$ is generated by $iT^i$ but the translation of the
coordinates $\theta_1^i,\theta_2^i$ with
$i=n+1,...,d$ is generated by $T^i$.
So let us define the generators of the $U(1)^d$ gauge (before
the rotation of the $\theta$ coordinates) by $\T^j=T^j$ for $j=1,
...,n$ and $\T^j=-iT^j$ for $j=n+1,...,d$. Then $\tr \T^i\T^j=\eta
 ^{ij}$. Now we want to repeat the procedure we were using in order
 to obtain the dual models.
Consider a rotation
$$\theta_1\rightarrow \theta_1'=O_1\theta_1\;\;\;;\;\;\;
\theta_2\rightarrow \theta_2'=O_2\theta_2\eqn\rota$$
The generators of the currents $J,\bar J$ in the rotated system
 are linear combinations of the generator
$\T^i$. A rotation ${\theta'_1}^i={O_1^i}_j\theta_1^j$ and
 ${\theta'_2}^i={O_2^i}_j\theta_2^j$ means that in the rotated action
 the corresponding $U(1)$ currents
 are generated by
  ${\T '}^i_L={O_1^i}_j\T^j$ and   ${\T '}^i_R={O_2^i}_j\T^j$.
Therefore in order to preserve the condition for anomaly
cancellation \bb \ the matrices $O_1$ and $O_2$ must be
$O(n,d-n)$ matrices (namely $O_1\eta O_1^T=O_2\eta O_2^T=\eta$).
Hence, when $d-n$ of the $U(1)$ isometries have opposite sign
the result \ggff \    for the inverse metric
of the dual models changes
as follows: The unit matrix multiplying ${\tilde c_G\over k}$ (the
${1\over k}$ correction to the inverse metric) is replaced by
$\eta$ and the matrices $O_1,O_2$  that generate the duality
are taken to be $O(n,d-n)$ matrices rather than $O(d)$.
In the case of dual models to the  ungauged WZW in section 5,
we can introduce
the $2d$ extra $U(1)$ currents with different signature. Suppose
we take $d-m$ pairs of $\varphi$ coordinates with negative signature
and the ungauged WZW has $d-n$ pairs of $\theta$ coordinates
which are non-compact, then the duality is obtained by
 $O(m+n,2d-m-n)$ matrices.

\chapter{\it Summary and Discussion}

In this work we have generalized the one loop $O(d,d)$
transformations to the exact to all orders case in WZW and WZW coset
models. A general $O(d,d)$ transformation can be decomposed to constant
coordinate transformations, a shift of the antisymmetric tensor by
a constant antisymmetric matrix and $O(d)\times O(d)$ transformations.
The first two are exact and in this paper
we have derived the exact $O(d)\times
O(d)$ transformations. We have found that for coset models $G/U(1)^d$
these transformations operate as follows: writing the inverse metric as
the semiclassical inverse metric plus the ${1\over k}$ corrections, only
the semiclassical part transforms- to a dual semiclassical inverse
metric. Therefore, although with the algebraic Hamiltonian approach
we could only derive the exact metric and dilaton field, knowing the
antisymmetric tensor to one loop order, we can find all the exact dual
metrics and dialton fields. In the ungauged WZW model with $2d$
abelian isometries we obtained the
dual metrics by considering the equivalent model $G\times U(1)^{2d}
/U(1)^{2d}$. Thus although  the ungauged WZW is exact, its dual models
receive ${1\over k}$ corrections with respect to the one loop order
transformation. Our analysis shows that
$\sqrt{G}e^{\Phi}$ is invariant
under the exact $O(d)\times O(d)$ transformations. In particular, in
the exact $G/U(1)^d$ coset models (as well as in the dual models)
$\sqrt{G}e^{\Phi}$ is independent of $k$.

Finally, we want to comment about the generality of our results to
a general exact CFT. It was shown by Tseytlin\Ref\tseytlin{A.A. Tseytlin
\journal Phys.Lett.&B288 (92) 279; \journal Phys.Rev.&D47 (93) 3421}
that there exist a $D+2$ dimensional background with a terget space
metric having a covariantly constant null Killing vector and a flat
'transverse' part that is exact to all orders.
The corresponding $\sigma$-
model is   invariant under D+1 abelian isometries.
In a recent work\Ref\tseyt{C. Klimcik and
A.A. Tseytlin, "Duality Invariant Class of Exact String Backgrounds",
preprint CERN-TH.7069/93} it was shown that a background of this type
is transformed by the one loop order $O(D,D)$ transformations
(only in the transverse directions)
to the same class of exact solutions.
This specific example requires a special analysis, but we speculate
that this  class  of exact solutions
can be obtained by gauging a larger
background followed by a redefining the fields so that the
${1\over k}$ correction is absorbed and disappears.
 Then the result in $\lbrack\tseyt\rbrack$ can be considered
as a particular example to our results.

\refout
\bye